Heating Up in NBA Free Throw Shooting
Paul R. Pudaite
January 12, 2018

Abstract

I demonstrate that repetition heats players up, while interruption cools players down in NBA free throw shooting.  My analysis also suggests that fatigue and stress come into play.  If, as seems likely, all four of these effects have comparable impact on field goal shooting, they would justify strategic choices throughout a basketball game that take into account the 'hot hand.'

More generally my analysis motivates approaching causal investigation of the variation in the quality of all types of human performance by seeking to operationalize and measure these effects.  Viewing the hot hand as a dynamic, causal process motivates an alternative application of the concept of the 'hot hand':  instead of trying to detect which player happens to be hot at the moment, promote that which heats up you and your allies.

Perspective on the Hot Hand

In much of the literature and in common lore, the 'hot hand' has been conceptualized (whether implicitly or explicitly) as a directly observable phenomenon.  Gelman (2015) states that "the hot hand effect is subtle to detect" but may still be found "… if you reanalyze the … data carefully".  With this in mind I develop mathematical models, employing Bayesian data analysis where required.

Furthermore, rather than viewing the hot hand as a discrete state, I propose investigating the quality of each individual human's performance as a dynamic process with continuous variation in magnitude.

Whence does the hot hand arise?  Is it a state that mysteriously appears and then vanishes?  My proposal is that 'hands' are continually heating up or cooling down, and what has been dubbed 'the hot hand' corresponds to an imprecise upper zone along a spectrum of performance.  Factors that improve performance *cause* entry into this zone; factors that degrade performance *cause* the exit or *prevent* its occurrence.

This perspective of the hot hand arising from a causal process, in contrast to an evanescent state, motivates a different method of application.  Rather than undertake the difficult task of identifying who's hot (whether by the offense to get him the ball, or the defense to shut him down), focus on fostering that which will cause you to heat up, and that which hinders your opponent's ability to do likewise.

Introduction

Variation in the quality of performance is ubiquitous in all human activities. I propose that *repetition*, *interruption*, *fatigue*, and *stress* make substantial contributions to this variation. Chang (2017) also presents evidence that error correction makes a contribution.

Performance generally improves with repetition, but eventually declines as fatigue sets in, and typically drops when interrupted or under stress. Individuals can also vary in their response to the outcome of their actions: for example, error correction produces improvement after failure.

Human activity can thus be viewed as a dynamic process subject to 'damping' (negative feedback from fatigue and error correction) and 'driving' (positive feedback from repetition and outcome reinforcement), along with perturbation (interruption). These effects are always present, but may not be easy to measure. In addition to noise from perturbation and intrinsic stochasticity, deterministic dynamics of damping and driving by themselves can generate chaos, in which high sensitivity to ongoing conditions makes prediction difficult.

I find strong evidence for the impact of repetition and interruption on NBA free throw shooting. I also present evidence suggestive of fatigue and stress, but further investigation will be required to be conclusive.

The Hot Hand in Free Throw Shooting

As noted by GVT (p. 304), free throws are "free from the contaminating effects of shot selection and opposing defense." This controlled situation for which hundreds of thousands of observations are now available makes it possible to accurately measure contributions from repetition and interruption, and also detect indications of fatigue and stress.

In Table 3 of "The Hot Hand in Basketball: On the Misperception of Random Sequences", Gilovich, Vallone and Tversky (1985; hencefort 'GVT') present "data for all pairs of free throws by Boston Celtics players during the 1980-1981 and the 1981-1982 seasons" (p. 304). GVT found "no evidence that the outcome of the second free throw is influenced by the outcome of the first free throw." We now know that this failure to find evidence arises in part from poor discrimination between 'hot' and 'cold' states. See, for example, Gelman (2015).

Ironically, GVT's Table 3 actually contains a wealth of information that unveils the path to effective exploration of the 'hot hand'. To begin with, it reveals repetition as a fundamental cause of 'heating up'. And it suggests that interruption can cause 'cooling down.'

Initial failures to observe variation in human performance suffered primarily from application of inadequate statistical tools (for example, Richardson 1945; GVT). By applying satisfactory statistical methodology, subsequent studies have demonstrated a 'hot hand' in, respectively, the outbreak of war (Houweling and Kuné 1984; Pudaite 1991) and in basketball field goal shooting (Bocskocsky, Ezekowitz and Stein 2014; Miller and Sanjurjo 2016).

The hot hand has been detected in free throw shooting by increasing sample size, hence the power of the statistical tests employed. Jeremy Arkes (2010) analyzed data for the entire NBA 2005-2006 season. Via fixed-effect logit modeling, he estimated that the difference in 2nd free throw percentage conditioned by the 1st free throw was *CD*=2.9%(0.8%) (standard error in parentheses).[1]

Preview of Key Evidence

Hot hand research typically employs some type of statistic that is conditioned by past outcomes. For pairs of free throws, the condition is limited to the binary outcome of the 1st shot. Gelman (2015) shows that even for a nontrivial effect size, conditioning on a single shot has a very large variance in comparison because past outcomes only weakly identify the shooter's state.

Fortuitously, free throw shooting exhibits a large causal effect that does not depend on past outcomes: *the act of shooting the first free throw, regardless of whether hit or missed* causes a typical player to hit 5- to 6-percentage-points higher on his 2nd shot than on his 1st. This effect is large enough (about twice the size of the conditional effect Arkes estimated) to emerge clearly in the following analysis performed on GVT's relatively small sample of free throw data:[2]

---

[1]However, Chang (2017) found that LeBron James's free throws in the 2016-2017 season exhibit error correction (a damping effect per my introduction above), hitting free throws
[2]GVT's table 3 includes enough data to precisely recover all of the 'raw' data, enabling the analysis presented in Table 1 above. (See Appendix 1 for details.)

| N | H1 | H2 | Pct1 | Pct2 | Pct2-Pct1 | StdErr | z |
|---|----|----|------|------|-----------|--------|---|
| 2049 | 1473 | 1590 | 71.9% | 77.6% | 5.7% | 1.4% | 4.21 |

| | |
|---|---|
| N: | Number of pairs of free throws |
| H1: | Number of 1st free throws hit |
| H2: | Number of 2nd free throws hit |
| Pct1: | Percentage of 1st free throws hit |
| Pct2: | Percentage of 1st free throws hit |
| StdErr: | Classical standard error of Pct2-Pct1 |
| z: | Classical standard score of Pct2-Pct1 |

Table 1[3]
Nine Members of the Boston Celtics during the 1980-1981 and 1981-1982 seasons

Observing this increased accuracy on the 2nd shot motivates the hypothesis that NBA players 'heat up' *during* trips to the free throw line.

Because players frequently receive more than one trip to the free throw line during a game, Table 1 also hints that players 'cool' down *between* trips to the free throw line. Tabulating fourteen seasons of NBA confirms this:

| Situation | N | H1 | H2 | Pct1 | Pct2 | Pct2-Pct1 | z |
|-----------|---|----|----|------|------|-----------|---|
| S1: first of 2+ Trips of 2+ Shots | 79,771 | 58,226 | 62,436 | 73.0% | 78.3% | 5.3% | 24.598 |
| S2: second of 2+ Trips of 2+ Shots | 79,771 | 59,176 | 62,411 | 74.2% | 78.2% | 4.1% | 19.043 |
| | | | Pctk[S2]-Pctk[S1]: | 1.2% | 0.0% | | |
| | | | Classical Standard Error: | 0.2% | 0.2% | | |
| | | | Classical Standard Score: | 5.395 | -0.152 | | |

Table 2
1233 players, NBA 2000-2001 through 2013-2014 seasons

Although the 1st shot of a NBA player's second trip to the line in a game repeats his action from his first free throw trip of the game, we see a substantial drop in success rate compared to the last shot of his first trip (74.2% vs 78.3%). But also observe that the players still exhibit some repetition benefit: their 1st shot percentage is higher on the second trip than on the first trip to the free throw line (74.2% vs 73.0%).

Table 2 conclusively establishes that *repetition* and *interruption* are capable of causing variation in human performance. But the demonstration is light on rigor because of my informal definitions of *repetition* as a sequence of free throw shots by one player during one trip to the free throw line, and *interruption* as whatever transpires between the player's trips to the free throw line.  It will be important for future research to investigate how best to operationalize these concepts in a variety of settings.

---

[3] Note on terminology: *percentage* refers to observed success rate; *probability* refers to the statistical expectation of success rate, which are not directly observable.

Free Throw Shooting Data

The pattern of 'heating up' in GVT's free throw data (Table 1 above) also appears in NBA play-by-play data for 2000-2001 through 2013-2014 seasons, with much greater statistical support:

| Situation | N | H1 | H2 | H3 | Pct1 | Pct2 | Pct3 | $\delta(1,2)$ | $\delta(2,3)$ | $Z(1,2)$ | $Z(2,3)$ |
|---|---|---|---|---|---|---|---|---|---|---|---|
| Exactly 1 | 80,940 | 59,039 | | | 72.9% | | | | | | |
| Exactly 2 | 382,031 | 279,703 | 297,207 | | 73.2% | 77.8% | | 4.6% | | 46.56 | |
| 3+ | 4,638 | 3,622 | 3,861 | 3,943 | 78.1% | 83.2% | 85.0% | 5.2% | 1.8% | 6.28 | 2.16 |
| Total | | | | | 73.2% | 77.9% | 85.0% | | | | |

| | |
|---|---|
| N: | Number of free throws in this situation |
| Hk: | Number of $k^{th}$ free throws hit |
| Pctk: | Percentage of $k^{th}$ free throws hit |
| $\delta(j,k)$: | Pctk-Pctj |
| $Z(j,k)$: | Classical standard score of Pctk-Pctj |

Table 3
Results of Single Trips to the Free Throw Line
NBA 2000-2001 through 2013-2014 seasons

In the additional "3+" row – single trips to the line for three or more free throws – performance continues to improve with additional *repetition*, at least within a single trip. To more dramatically frame the overall improvement during a trip to the free throw line: when NBA players went to the line for three or more free throws, they missed 46% more (1016 vs. 695) of their 1st than 3rd attempts!

The 'improvement' from 2nd to 3rd free throw appears even greater in the 'Total' row (7.1 percentage-points, from 77.9% to 85.0%).  However, this level of data aggregation greatly overestimates the improvement of *individual players* from 2nd to 3rd free throw success rates.  Trips for three or more shots occur when a player is fouled attempting a three-point field goal, or when a two shot foul is compounded with a technical foul.  As result poorer shooters receive far fewer of these opportunities.

In general, more detailed classical analysis becomes weaker because adding conditions (1) subdivides the data into geometrically smaller subsamples (aka 'bins'), and (2) may introduce new confounders.  Attempting to control for potential confounders via binning reduces sample size still further.  Applying Bayesian methods with the player as the unit of analysis (rather than the league) is more effective.

Model 1: Bayesian analysis of individual player data

Bayesian data analysis provides superior statistical control by permitting a model to incorporate otherwise confounding variables that are difficult or impossible to estimate classically – in this case, individual free throw shooting ability, which varies substantially across the players in the NBA. Explicit modeling of individual ability enables more accurate estimates of the impacts of repetition and interruption, and reveals possible effects of fatigue and stress.

I organized Model 1 as a sequence of four nested components. Starting from the inside out:

$$\{\mathbf{Y}_{ij}\}_{j=1}^{N_i} : \text{player } i \text{ free throw trip } j \text{ outcomes}$$

$$N_i : \text{number of free throw trips by player } i$$

$$\mathbf{Y}_{ij} = (Y_{ijk})_{k=1}^{n_{ij}}$$

$$n_{ij} : \text{number of free throws in trip } j \text{ by player } i$$

Model 1, component 1: sample data[4]

Making the simplifying assumption that a player's probability of making a free throw does not depend on the outcome of previous free throws in the trip to the line, we can write:

$$\mathbf{P}_{ij} = (P_{ijk})_{k=1}^{n_{ij}}$$

$$Y_{ijk} = B(1, P_{ijk})$$

$$B(n, p) : \text{binomial random variable for } n \text{ trials with probability } p$$

Model 1, component 2: binomial distribution

Although Arkes (2014) and Chang (2017) have reported evidence that 1st free throw outcomes affect 2nd free throw percentage, we will nonetheless see that Model 1 provides a satisfactory account of the impact of conditioning the 2nd free throw on the outcome of the 1st (see Appendix 3).

---

[4]Unless otherwise noted, upper case roman letters indicate random scalar variables, and bold indicates random vectors.

To acknowledge the variability of performance within each individual player, assume that $\mathbf{P}_{ij}$ is drawn from a logistically transformed multivariate normally distributed random variable. As a further simplifying assumption, assume that the mean and variance of these random variables do not depend on the total number of free throws in a trip to the line. We can then write:

$$P_{ijk} = \frac{e^{X_{ijk}}}{e^{X_{ijk}} + 1}$$

$$= f(X_{ijk})$$

$$\mathbf{X}_{ij} = N(\mu_i, \Sigma_i)$$

$$\mu_i \in \Re^n$$

$$\Sigma_i \in \Re^{n \times n}$$

$n$: maximum number of free throws in one trip to the line

Model 1, component 3: logistic distribution
(*Intra*-individual variability)

Because fewer than 1% of the trips to the line are for three or more free throws (4638 out of 467,609), I reduced the computational requirements by estimating the distribution of the moments only for the 1st and 2nd free throws.

Estimation-maximization of this model for the NBA data produces $\Psi_1$, a discrete distribution of (model 1) *profiles*, i.e., hypotheses over $(\mu_i, \Sigma_i)$, $\mu_i \in \Re^2, \Sigma_i \in \Re^{2 \times 2}$:

$$(\mu_i, \Sigma_i): \text{drawn from } \Psi_1$$

$$\Psi_1 = \{(\pi_{1m}, \mu_{1m}, \Sigma_{1m})\}_{m=1}^{M_1}$$

$$\mu_{1m} \in \Re^2$$

$$\Sigma_{1m} \in \Re^{2 \times 2}$$

$$\pi_{1m} = \Pr[(\mu_i, \Sigma_i) = (\mu_{1m}, \Sigma_{1m}) | \Psi_1]$$

Model 1, component 4: hierarchical Bayes
(*Inter*-individual variability)

Figure 1 depicts Model 1 as a probabilistic graphical model:

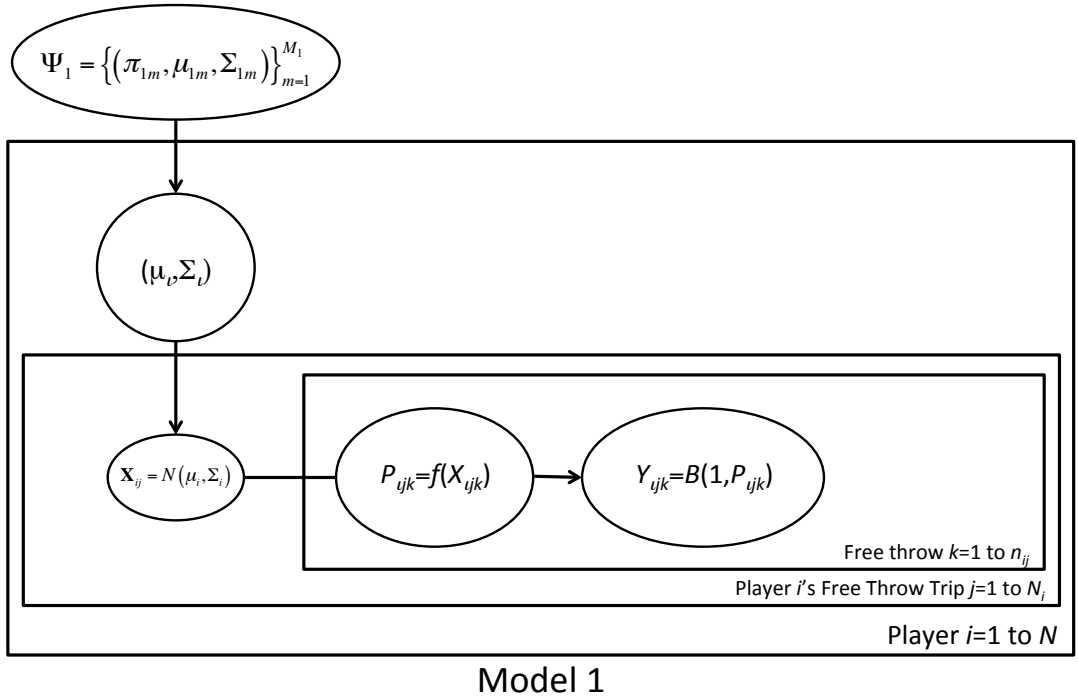

Model 1

Figure 1

The lines with arrows indicate random draws. Plates (the rectangular regions) indicate independent, identically distributed draws. Lines without arrows indicate deterministic relationships.

Figures 2a and 2b show 84%[5] confidence regions for each of the $M_1=56$ profiles present in the support of $\Psi_1$; figure 2b depicts the profile's prior probability in the vertical axis.

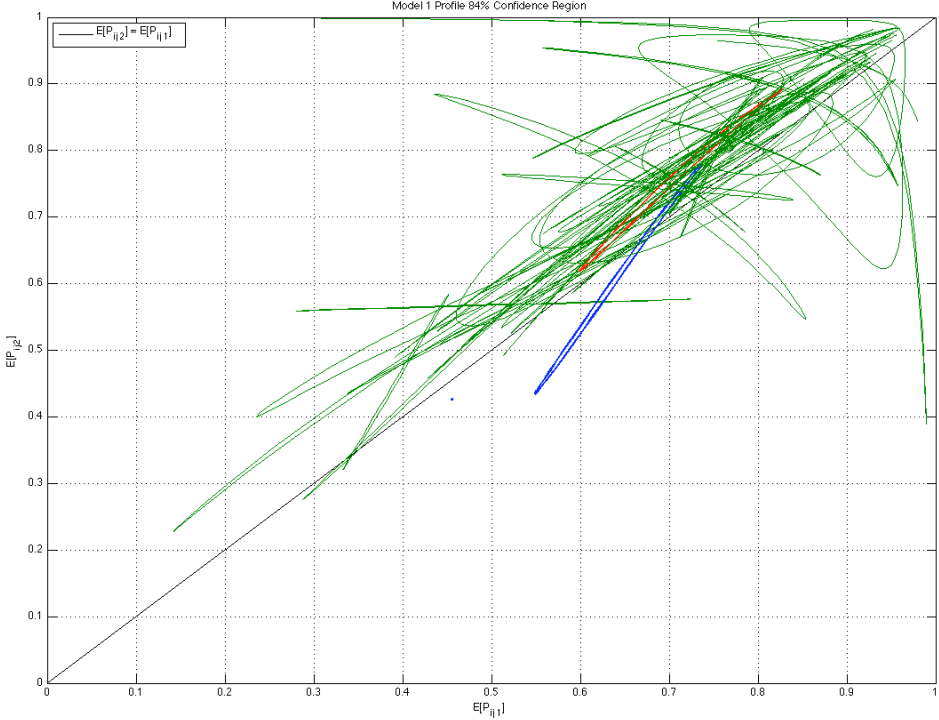

Figure 2a

---

[5] I chose 84% because most of the confidence regions are so eccentric that they are close to linear (in logit space). In one dimension, an 84% confidence region extends one standard deviation in each direction.

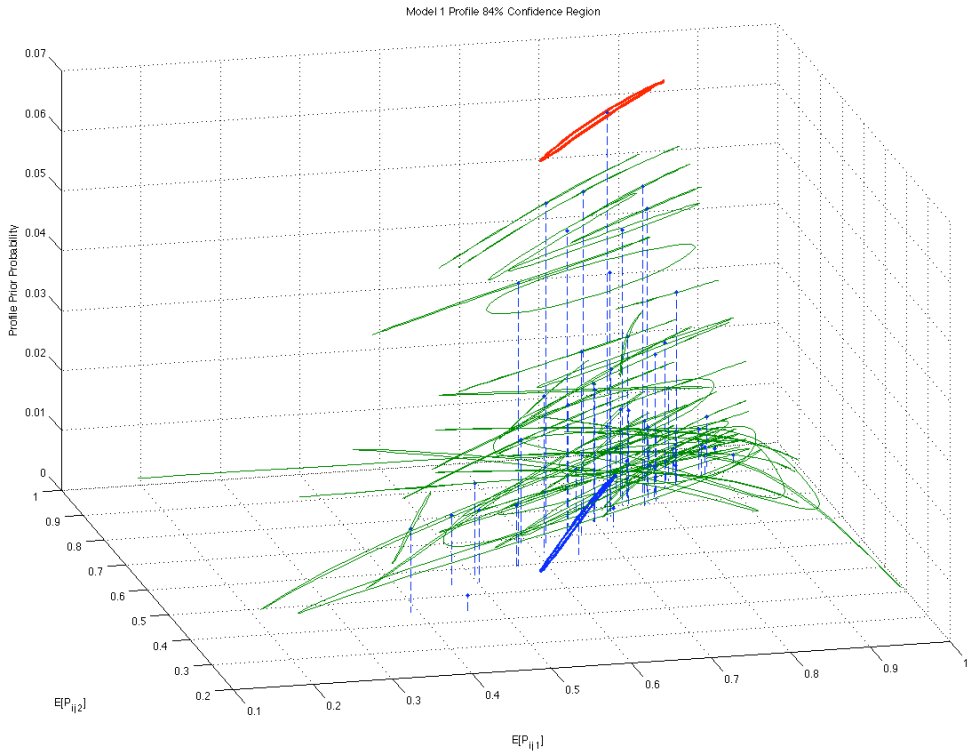

Figure 2b

In figure 2b, the vertical blue dashed stems help locate each profile's mean value. In each figure, the thick red curve highlights $\theta_1$, the modal (i.e., highest prior probability) profile.

The performance variation we observe for each profile is due to *intra*-individual variability (of a player's ability to shoot free throws) across trips to the free throw line for a player's observed career within the data. This is explained further in Appendix 2 using the example of LeBron James (LeBron has the highest posterior probability of belonging to $\theta_1$).

Players vary in their response to free throw repetition within a single trip to the line. Most of the profiles lie roughly parallel to the diagonal axis, $E[P_{ij2}|\theta_m]= E[P_{ij1}|\theta_m]$. But there are also many profiles whose confidence regions extend below the diagonal: for *some* trips to the line by such players, repetition has the atypical effect of *worsening* performance.

There are even two profiles for which $\mu_{1m2} < \mu_{1m1}$, i.e., the player shoots 2nd free throws worse than 1st free throws *most* of the time. Thick blue curves highlight these profiles in both figures (one is highly eccentric, the other resolves as a point). Not many NBA players follow these profiles; $\Pr[\mu_{1m2} < \mu_{1m1}|\Psi_1]$=0.0090.

Figure 3 shows posterior profile estimates for LeBron James conditional on the outcome of his 1st shot in a trip to the free throw line:

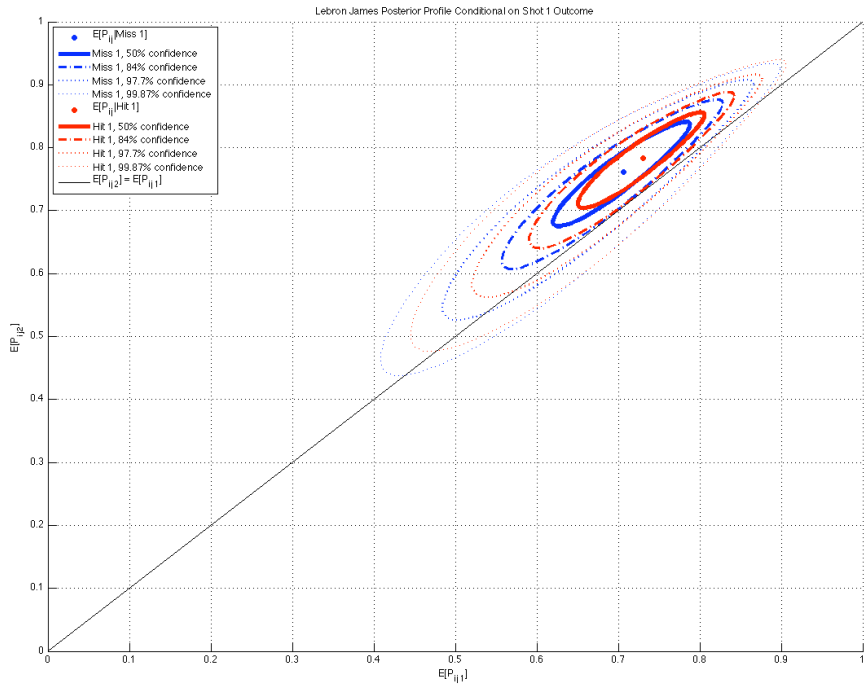

Figure 3

The red and blue points, located at (73.0%, 78.3%) and (70.6%, 76.1%), respectively, show LeBron's conditional expected values.  We expect LeBron to hit 2nd free throws 2.2 percentage-points more often if he makes his 1st than if he misses (78.3%-76.1%).

Figure 3 illustrates Gelman's discussion of the difficulty of identifying a 'hot hand,' showing that GVT could have detected positive serial correlation with sufficient data.[6]  Figure 3 also provides an example of the way in which Model 1 accurately accounts for Arkes's result on the effect of the 1st free throw on the 2nd.

See Appendix 3 for further discussion of serial correlation and the difference in 2nd free throw percentage conditioned on 1st free throw outcome.  Model 1 accounts for the empirical values of all of these statistics.  Appendix 3 also introduces an intriguing anomaly for future investigation.

---

[6]For example if we set a null hypothesis test threshold at a standard score of $z=2$, given Lebron's posterior profile estimate, it would take $N=1487$ trips to the free throw line to generate 50% power for that test.

Model 2: Intra-game performance variation for each trip to the line

Model 2 uses $\Psi_1$, the profile distribution estimated for Model 1. The new component in Model 2 is $\Delta_h$, the displacement from a player's Model 1 profile as a function of $h$, the 'intra-game trip index' of that player. For each value of $h$, $\Delta_h$ is drawn once for all of the data from $\Psi_2$, the prior distribution of these displacements.[7]

$$P_{ijk} = \frac{e^{Z_{ijk}}}{e^{Z_{ijk}} + 1}$$

$$\mathbf{Z}_{ij} = \left(Z_{ijk}\right)_{k=1}^{2}$$

$$\mathbf{Z}_{ij} = \mathbf{X}_{ij} + \Delta_{h(i,j)}$$

$h(i,j)$: index of trip to the line within game for player $i$'s career trip $j$

$$\Delta_h = \left(\Delta_{hk}\right)_{k=1}^{2}$$
$\qquad$: drawn from $\Psi_2$

$$\Psi_2 = N\left(\begin{pmatrix} 0 \\ 0 \end{pmatrix}, \Sigma_\Delta\right)$$

$$\Sigma_\Delta \in \Re^{2\times 2}$$

$$\mathbf{X}_{ij} = N\left(\mu_i, \Sigma_i\right)$$

$\left(\mu_i, \Sigma_i\right)$: drawn from $\Psi_1$

---

[7] Adding inter-player variability would provide better sampling control, but model 2 provides satisfactory results for this paper. Future research will likely benefit from incorporating this hierarchical level.

Figure 4 depicts Model 2 as a probabilistic graphical model:

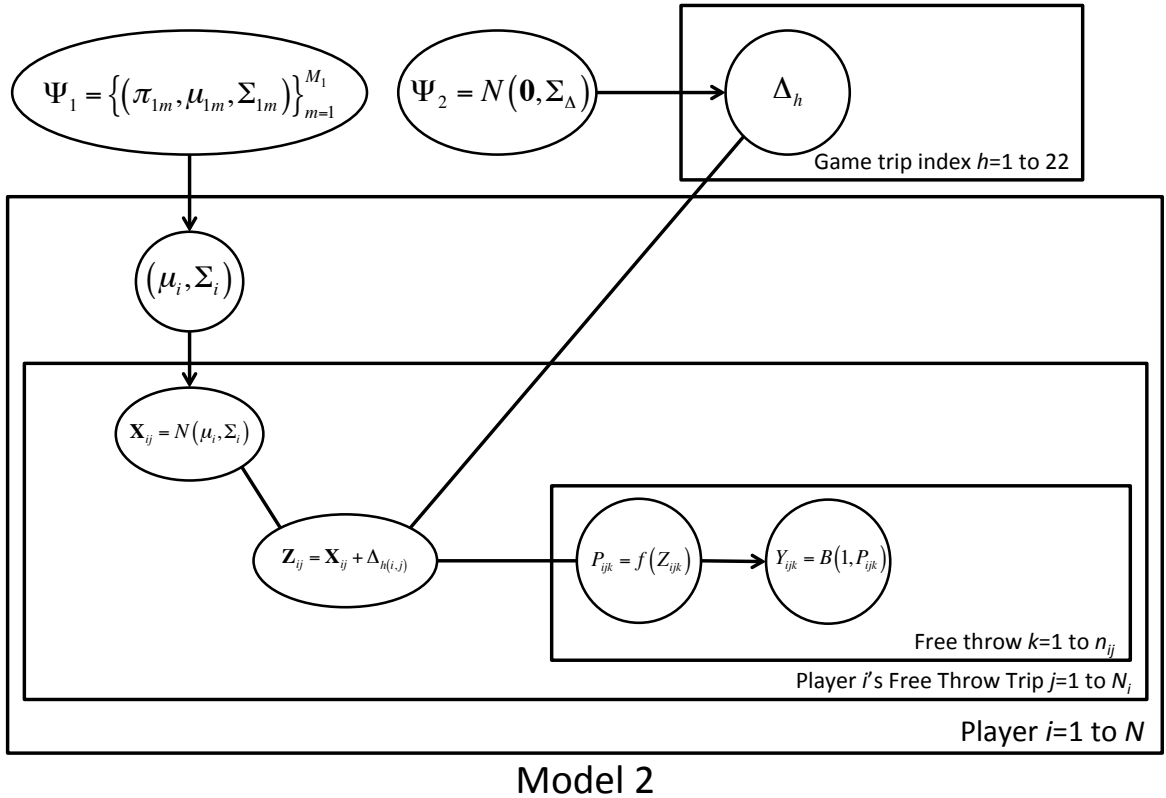

Model 2

Figure 4

Via expectation maximization I obtained $\Sigma_\Delta = \begin{pmatrix} 0.0402 & 0.0080 \\ 0.0080 & 0.0346 \end{pmatrix}$; estimates of ($\Delta_h$) appear in the next chart. $\Sigma_\Delta$ corresponds to a standard deviation of 0.20 logit units for 1st free throws and 0.19 units for 2nd free throws, with a correlation of $\rho = 0.21$. For a typical NBA player (75.4% free throw percentage), this transforms to a 3.5 percentage point standard deviation in expected 1st and 2nd free throw percentage across trips to the line.

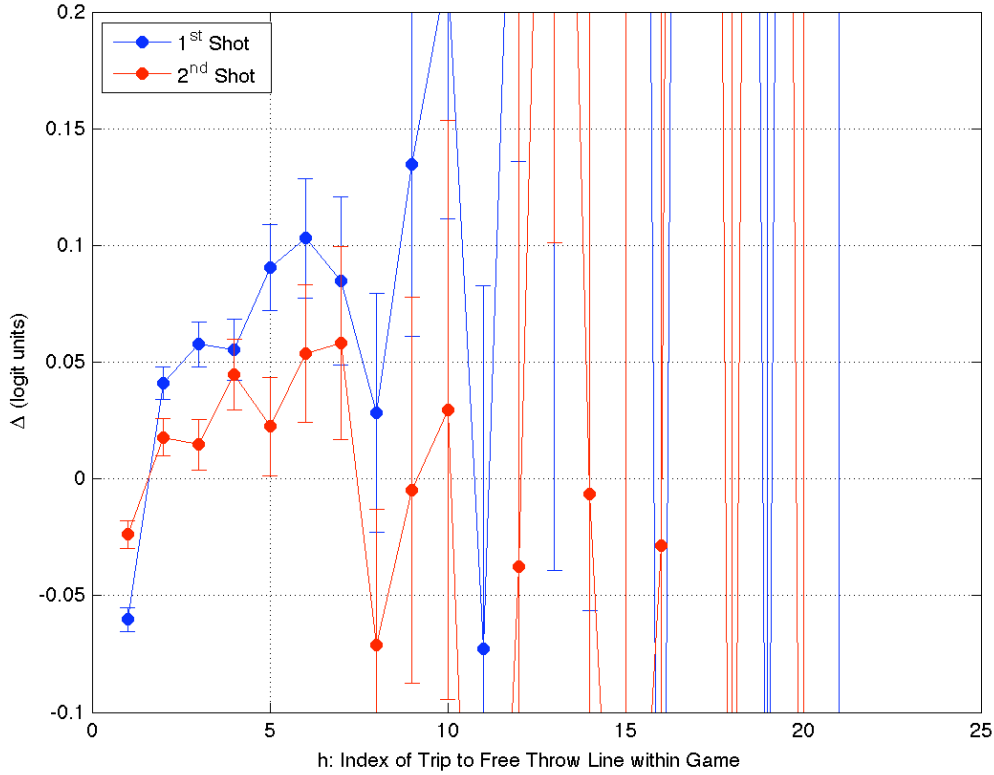

Figure 5

We see in figure 5 that players definitely improve on their second trip to the free throw line in a game, compared to their first trip, $\chi^2_{2,\Delta(2^{nd}\,trip)-\Delta(1^{st}\,trip)} = 12.378$.[8]

The improvement appears to continue through the sixth or seventh trip to the line, and there may be a decline for the eighth and subsequent trips to the line. However, the sample size declines geometrically as $h$ increases. In addition, as mentioned in footnote 7 (above), Model 2 ignores the variation across individual players. As $h$ increases, $(\Delta_h)$ is estimated from an ever smaller, less representative subset of players. As a result, the measurement errors are too large to make these claims (continuing improvement from the second through sixth or seventh trip, and subsequent decline) with confidence.

---

[8] The Mahalanobis distance for $k$-dimensional random vector $W$, $\chi^2_{k,W} = \sqrt{W'Var[W]^{-1}W}$, follows a $\chi^2$ distribution with $k$ degrees of freedom.

Model 3: Intra-game performance as a function of game time elapsed.

Extending Model 2 to permit individual variation of $(\Delta_{hk})$ could help alleviate the sampling bias, but can't overcome the declining sample size with respect to *h*. To better explore the possible role of fatigue in free throw shooting, I estimated the following model:

$$\mathbf{Z}_{ij} = \mathbf{X}_{ij} + \Delta_{\hat{h}(i,j)}(t_{ij})$$

$$\hat{h}(i,j) = \min(h(i,j), 2)$$

$\quad t_{ij}$ : game time elapsed in player *i*'s career trip *j*

$$\Delta_h(t) = \left(\Delta_{hk}(t)\right)_{k=1}^{2}$$

$\quad$ : drawn from $\Psi_2$

Figure 6 depicts Model 3 as a probabilistic graphical model:

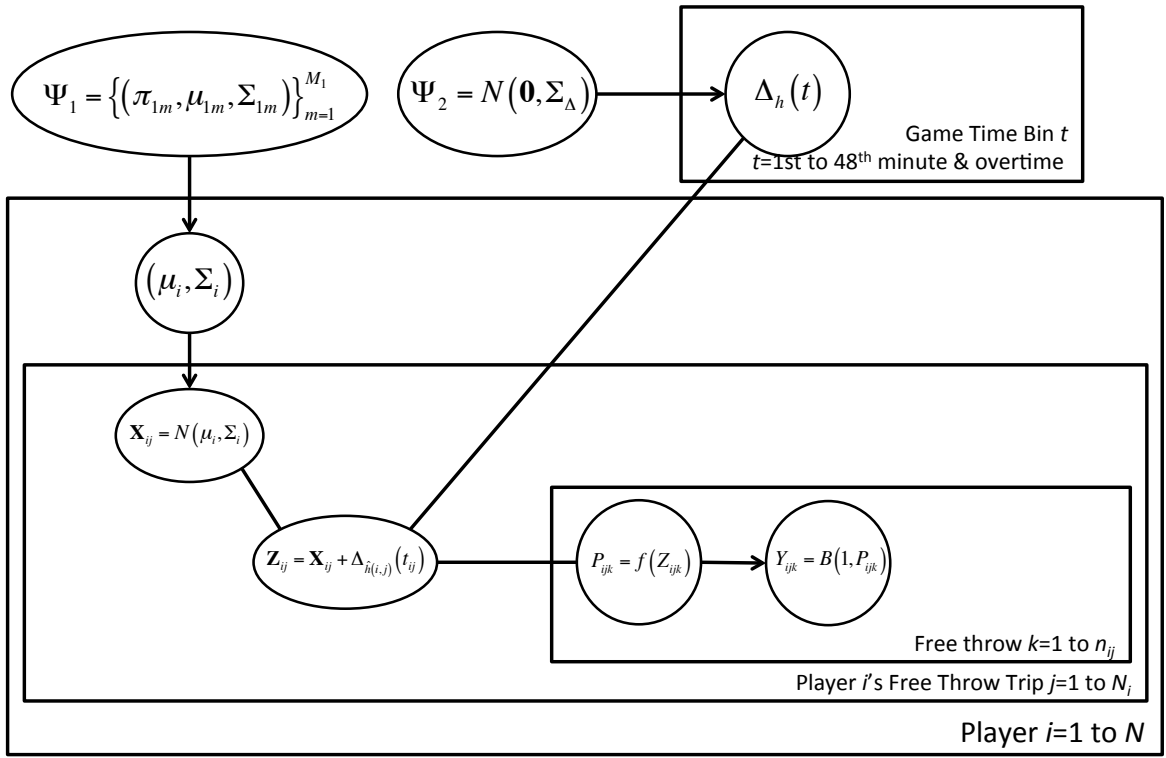

Model 3

Figure 6

Model 3 incorporates the two best supported findings from Model 2: (1) the first trip to the free throw line differs substantially from subsequent trips, and (2) $\Psi_2$, the distribution of intra-game free throw shooting displacements from the player's career profile estimated via expectation-maximization. Model 3 introduces no new parameters, hence no additional degrees of freedom.

I binned observations within regulation by the minute, and collected all overtime observations into a final 49th bin. Figures 7a and 7b show $\hat{\Delta}_h(t)$, binned estimates, and $\tilde{\Delta}_h(t)$, Kalman filter estimates for, respectively, first and subsequent trips to the free throw line. Overtime estimates are plotted somewhat arbitrarily at $t$ = 50.5 minutes, the midpoint of the first overtime.

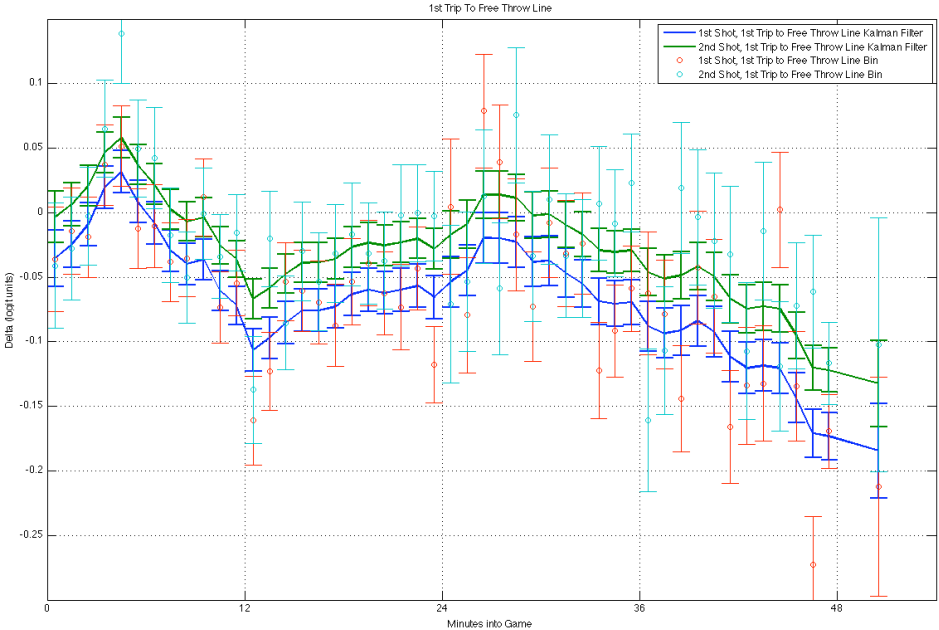

Figure 7a

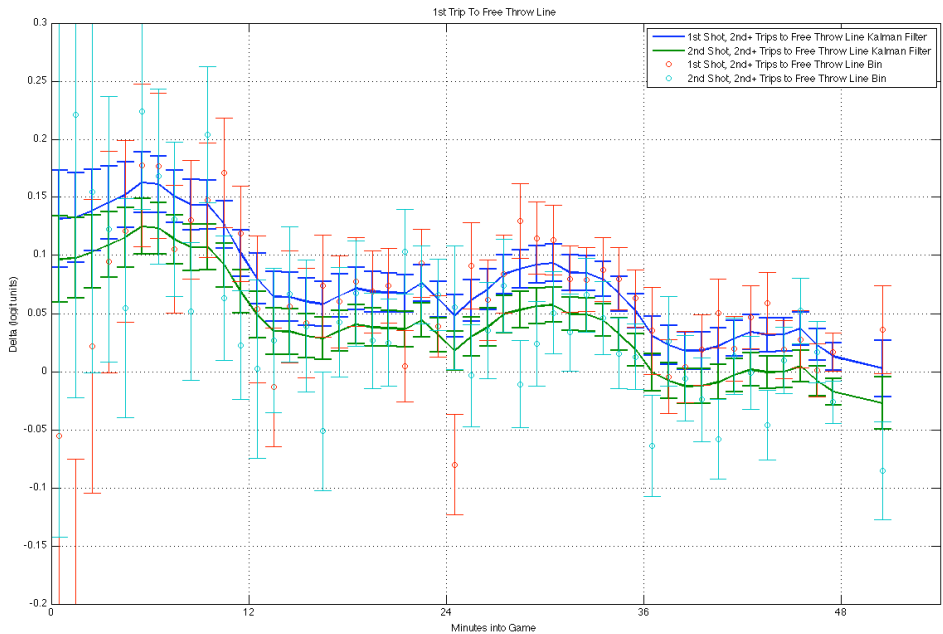

Figure 7b

$\tilde{\Delta}_1(t)$ and $\tilde{\Delta}_2(t)$ both decline over the course of the game, providing some support for the hypothesis that fatigue hampers performance. Table 4 presents Mahanalobis distances for these trends and additional sub-trends visible in the charts.

| Trend    | B0,1 | B1,1 | B0,2 | B1,2 | $\chi^2_{2,1}$ | $\chi^2_{2,2}$ |
|----------|------|------|------|------|----------------|----------------|
| Decrease | 1    | 49   | 1    | 49   | 3.525          | 2.864          |
| Increase | 1    | 5    | 1    | 7    | 2.527          | 0.624          |
| Decrease | 5    | 13   | 7    | 25   | 6.021          | 3.849          |
| Increase | 13   | 27   | 25   | 31   | 3.465          | 3.849          |
| Decrease | 27   | 49   | 31   | 49   | 3.999          | 1.905          |
| Decrease |      |      | 31   | 39   |                | 3.420          |
| Increase |      |      | 39   | 46   |                | 0.903          |
| Decrease |      |      | 46   | 49   |                | 1.259          |

B0,h:   Index of starting bin for trend in $\tilde{\Delta}_h(t)$
B1,h:   Index of ending bin for trend in $\tilde{\Delta}_h(t)$
$\chi^2_{2,h}$:   Mahalanobis distance for trend in $\tilde{\Delta}_h(t)$

Table 4
Trend Statistics within $\tilde{\Delta}_h(t)$
NBA 2000-2001 through 2013-2014 seasons

Other factors could come into play. For example, it seems plausible that there is more stress the first time a player goes to the free throw line the deeper into the game that occurs; this could contribute to the steeper decline of $\tilde{\Delta}_1(t)$ compared to $\tilde{\Delta}_2(t)$.

Visually, there are four similar sub-trends shared by the two charts.[9] In each case, the direction changes for $\tilde{\Delta}_2(t)$ further into the game than for $\tilde{\Delta}_1(t)$. The similarities encourage further investigation. What causes the improving performance at the start and middle of the game? Fatigue seems to be a plausible explanation for the two declining trends, but why do they occur earlier for the first trip to the line? Patterns of player substitution may play a role, which would entail incorporating information on how much time each player has been in the game at each trip to the free throw line.

One of the steeper drops occurs at the end of the game, from the 45 minute bin through overtime. Stress seems particularly likely to play a role at the end of the game, particularly if the score is close. It may be possible to isolate this aspect of stress by adding a suitable explanatory variable, such as the change in the probability of the player's team winning conditional on the outcome of his trip to the line. We would certainly expect this to vary across players; in particular, this analysis could identify 'clutch' players: those who perform better under stress than their career profile.

Some Implications for Basketball Strategy

Coaches have some ability to harness the benefits of repetition on offense and interruption on defense.

1. If a technical is called in conjunction with a common foul, there may often be situations in which the team should choose to have the common-fouled team member shoot three free throws, rather than switching to a 'superior' free throw shooter for the technical, particularly if the otherwise superior shooter would be making his first free throw attempt of the game. The strategic alternatives boil down to anticipated success rates of the 'superior' shooter's 1st shot vs. the fouled shooter's 3rd shot. Typically a 6 to 8 percentage-point adjustment should be made, and may often be even larger.

2. We know that coaches sometimes attempt to 'ice' a free throw shooter by calling a timeout. It remains to be seen whether the efficacy of this interruption strategy can be evaluated from available data.

3. The opposing team can also interrupt a foul shooter by substituting in a player between his 1st and last free throws in a trip to the line. Anecdotally, I have observed many NBA coaches doing this, while NCAA coaches typically make their substitutions before free throws begin, or after the shooter hits his last free throw (substitution is not permitted if the last free throw is missed).

---

[9] Keep in mind that the Mahalanobis distances of these intra-game trends are inflated due to post hoc selection of the peak and valley times.

Speculative Strategy Implications

There are likely to be even more valuable strategic opportunities for employing repetition and interruption in field goal shooting than for free throws.  Optimal deployment will require accurate identification and measurement of these and other causal and potentially strategic factors in field goal shooting.  Bocskocsky, Ezekowitz, and Stein (2014) have now performed the more difficult task (at least compared to free throws) of identifying hot field goal shooters.  Their research may help guide causal study.

4.  When a player 'bricks' a field goal attempt in a pickup basketball game, his teammates may never pass him the ball again during the game.  Because of (a) the wide range of ability in such games and (b) the lack of familiarity among some teammates, this reaction may be justified by intuitive Bayesian inference.  However, in the NBA, the coach should have an accurate appraisal of the abilities of each of his players.  He should make at most small adjustments to this appraisal in response to outcomes within a single game (recall LeBron's conditional posteriors in Figure 3), and he needs to ensure that his players also understand this.

5.  As with free throws, the mere act of field goal shooting may *improve* a player's probability of hitting his next field goal, provided it is sufficiently similar in location and time.  This effect needs to be accurately measured, if indeed it even exists.  Anecdotally, Phil Jackson's championship Chicago Bulls teams were the first I saw behave in accord with this principle, viz., going right back to a player after a miss provided that he had a similar opportunity on the next possession.

6.  Conversely, prevent opponents from getting a sequence of similar shooting opportunities.  Don't necessarily increase defensive pressure on players who have made field goals.  Instead, force opponents to take a sufficiently different shot, regardless of previous outcome.

7.  The 'heat check' is likely ill-advised unless it is sufficiently similar to previous shots.  Again, 'sufficient similarity' remains to be quantified.

Possibilities for Future Research

Model 1 can be adapted to measure intra-player performance variation at different time scales.  For individual players, estimate intra-game variance, single season variance, and career variance.

Study intra-player performance as a time series. Over what time scales can we identify 'hot' vs 'cold' streaks? Can we learn from this how players can end cold streaks, and lengthen hot streaks?

Can clutch performance be trained? Is there a way to help players avoid choking?

Fatigue may be reduced when play stops due to fouls, timeouts, etc. Ideally, we would want to add the time of day as an additional explanatory variable, but this may be hard to obtain.

Distinguish between physical and mental fatigue. Mental fatigue has numerous aspects, including boredom and over-stimulation.

Improve the implementation of the models in this paper. All of the components can be estimated simultaneously (possibly increasing the risk of overfitting). Incorporate individual variation in the impact of game time.

Comments and ideas welcome!

Appendix 1: Raw Data Recovery from GVT Table 3

Table 3 in GVT (p. 305) reports the observed percentages of hitting a second free throw conditioned on each outcome of the first free throw, along with the number of shots taken in each condition, and the (normalized) serial correlation for nine members of the Boston Celtics during the 1980-1981 and 1981-1982 seasons. The small sample sizes make it possible to unambiguously determine the integer number of shots made in each condition, enabling full recovery of the 'raw' data:

| Name | N | MM | MH | HM | HH |
|---|---|---|---|---|---|
| Larry Bird | 338 | 5 | 48 | 34 | 250 |
| Cedric Maxwell | 430 | 31 | 97 | 57 | 245 |
| Robert Parish | 318 | 29 | 76 | 49 | 165 |
| Nate Archibald | 321 | 14 | 62 | 42 | 203 |
| Chris Ford | 73 | 5 | 17 | 15 | 36 |
| Kevin McHale | 177 | 20 | 29 | 35 | 93 |
| M. L. Carr | 83 | 5 | 21 | 18 | 39 |
| Rick Robey | 171 | 31 | 49 | 37 | 54 |
| Gerald Henderson | 138 | 8 | 29 | 24 | 77 |
| | | | | | |
| N: | Number of pairs of free throws | | | | |
| MM: | Miss $1^{st}$, Miss $2^{nd}$ | | | | |
| MH: | Miss $1^{st}$, Hit $2^{nd}$ | | | | |
| HM: | Hit $1^{st}$, Miss $2^{nd}$ | | | | |
| HH: | Hit $1^{st}$, Hit $2^{nd}$ | | | | |

Table A1
Nine Members, Boston Celtics, 1980-1 and 1981-2 seasons

From the raw data, we can obtain success rates for 1st free throws and 2nd free throws:

| Name | N | H1 | H2 | Pct1 | Pct2 | Pct2-Pct1 | StdErr | z |
|---|---|---|---|---|---|---|---|---|
| Bird | 338 | 285 | 298 | 84.3% | 88.2% | 3.9% | 2.6% | 1.57 |
| Maxwell | 430 | 302 | 342 | 70.2% | 79.5% | 9.3% | 3.0% | 3.15 |
| Parish | 318 | 213 | 241 | 67.0% | 75.8% | 8.8% | 3.6% | 2.36 |
| Archibald | 321 | 245 | 265 | 76.3% | 82.6% | 6.2% | 3.2% | 1.95 |
| Ford | 73 | 51 | 53 | 69.9% | 72.6% | 2.7% | 7.5% | 0.37 |
| McHale | 177 | 128 | 122 | 72.3% | 68.9% | -3.4% | 4.8% | -0.70 |
| Carr | 83 | 57 | 60 | 68.7% | 72.3% | 3.6% | 7.1% | 0.51 |
| Robey | 171 | 91 | 103 | 53.2% | 60.2% | 7.0% | 5.4% | 1.31 |
| Henderson | 138 | 101 | 106 | 73.2% | 76.8% | 3.6% | 5.2% | 0.70 |
|  |  |  |  |  |  |  |  |  |
| Total | 2049 | 1473 | 1590 | 71.9% | 77.6% | 5.7% | 1.4% | 4.21 |

| | |
|---|---|
| N: | Number of pairs of free throws |
| H1: | Number of 1st free throws hit |
| H2: | Number of 2nd free throws hit |
| Pct1: | Percentage of 1st free throws hit |
| Pct2: | Percentage of 2nd free throws hit |
| StdErr: | Classical standard error of Pct2-Pct1 |
| z: | Standard score |

Table A2
Nine Members, Boston Celtics, 1980-1 and 1981-2 seasons

Appendix 2: Model 1 Profile Interpretation

The thick red curve centered at (72.8%, 78.5%, 6.6%) corresponds to $\theta_1$, the most common profile of NBA free throw shooters. In expectation, about 7% of the players in the data (81.8 of 1233) fall into this profile. The curve is highly eccentric, indicating that 1st and 2nd free throw probabilities co-vary tightly.

But the curve also covers a wide range: (59.9%,61.9%) at the lower end of the 84% confidence region to (82.8%,89.2%) at the upper end. On a trip to the line for two free throws, these players miss both 15% of time when 'cold', but less than 2% of the time when 'hot.' … And these players are more consistent than most!

From the perspective of free throw percentage, these players benefit more from repetition on the 'hot' end of the confidence region, with a 6.4% increase in free throw percentage from 1st to 2nd compared to a 2.0% increase in the 'cold' end. In logit space, the repetition benefit is even larger: +0.08 units when cold (0.40 to 0.48), +0.54 units (1.57 to 2.11) when hot.

Table A3 shows the five players with the highest $\theta_1$ profile posterior probability:[10]

| Player | Pr[$\theta_1$] | Seasons |
|---|---|---|
| L. James | 0.668 | 2003-2004 to 2013-2014 |
| T. McGrady | 0.630 | 2000-2001 to 2011-2012 |
| P. Gasol | 0.626 | 2001-2002 to 2013-2014 |
| T. Parker | 0.610 | 2001-2002 to 2013-2014 |
| E. Brand | 0.576 | 2000-2001 to 2013-2014 |

Table A3

Figure A1 provides more detail on $\Psi_1|\text{Lebron}$, LeBron's posterior profile estimate:

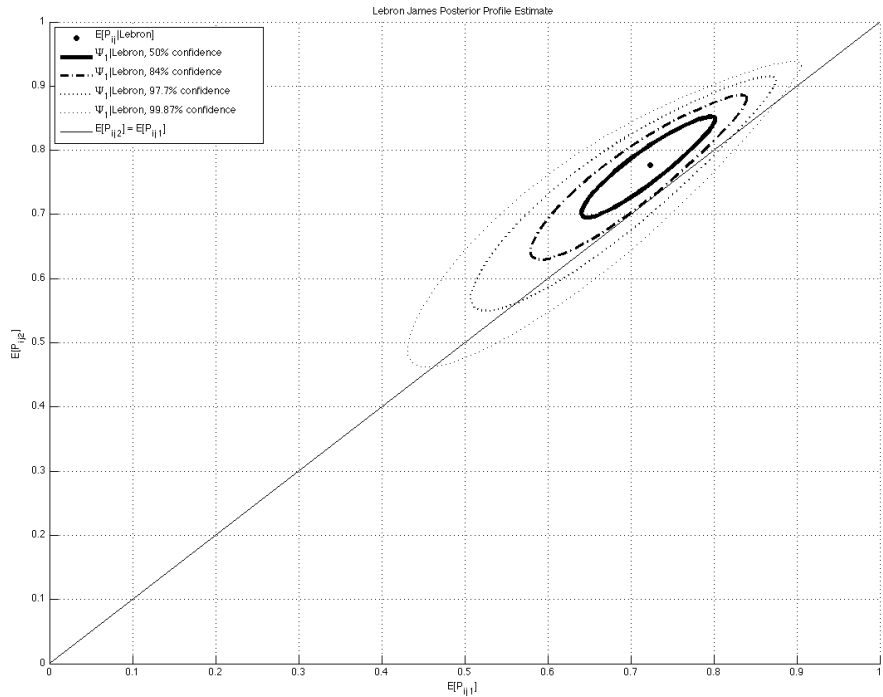

Figure A1

---

[10]Notational humor: $\arg\max \Pr\left[\theta_1 \Big| \{\mathbf{Y}_{ij}\}_{j=1}^{N_i}\right] = \text{Lebron James}$.

$\Psi_1|$Lebron combines measurement error (uncertainty in identifying LeBron's specific profile) and the actual variability in his free throw shooting across his eleven seasons in the data. Because of the prevalence of highly eccentric profiles in $\Psi_1$ that are roughly parallel to y=x, we can infer that the major axes of the confidence regions in Figure A1 primarily represents his actual variability, while measurement error lengthens the minor axes and reduces eccentricity of $\Psi_1|$Lebron compared to $\theta_1$.

Since LeBron has made thousands of trips to the free throw line, $\Psi_1|$Lebron asserts that it is highly likely that he has had many 'cold' trips to the line (say, $E[P_{Lebron,jk}] < 0.60$), and on the other extreme, comparably many trips 'in the zone' (say, $E[P_{Lebron,jk}] > 0.85$).

Appendix 3: Correlation Statistics

The following statistic is an unbiased estimator of the covariance of the probabilities of making the first two free throws in a trip to the line:

$\hat{R}_i$ : unbiased estimator of serial correlation of player $i$'s first two free throws

$$= \frac{1}{N_{i2}-1} \sum_{n_{ij} \geq 2} (Y_{ij1} - \bar{Y}_{i1})(Y_{ij2} - \bar{Y}_{i2})$$

$$\bar{Y}_{ik} = \sum_{j:n_{ij} \geq 2} Y_{ijk} \Big/ N_{i2}$$

$$N_{ih} = \sum_{j:n_{ij} \geq h} 1$$

$$E[\hat{R}_i] = Cov[P_{ij1}, P_{ij2}]$$

$$= \sigma_{12}$$

$$Var[\hat{R}_i] = \frac{P_{ij1}(1-P_{ij1})P_{ij2}(1-P_{ij2})}{N}$$

Serial correlation and conditional difference are effectively scaled versions of $\hat{R}_i$:

$R_i$ : (normalized) serial correlation of player $i$'s first two free throws

$$= \frac{1}{N_{i2} s_{i1} s_{i2}} \sum_{j:n_{ij} \geq 2} (Y_{ij1} - \bar{Y}_{i1})(Y_{ij2} - \bar{Y}_{i2})$$

$$E[R_i] \approx \frac{N_{i2} \sigma_{12}}{(N_{i2} - 1)\sqrt{E[P_{i1}](1 - E[P_{i1}])E[P_{i2}](1 - E[P_{i2}])}}$$

$CD_i$ : difference of player $i$'s 2nd free throw percentage, conditional on 1st free throw outcome

$$N_{iqr} = \frac{\sum_{j:n_{ij} \geq 2, Y_{ij1}=1} Y_{ij2}}{\sum_{j:n_{ij} \geq 2, Y_{ij1}=1} 1} - \frac{\sum_{j:n_{ij} \geq 2, Y_{ij1}=0} Y_{ij2}}{\sum_{j:n_{ij} \geq 2, Y_{ij1}=0} 1}$$

$$E[CD_i] \approx \frac{\sigma_{12}}{E[P_{i1}]}$$

The expected value formulas are approximate due to the presence of random variables in their denominators.

$\hat{R}_i$ best isolates the primary quantity of interest, covariation of expected 1st and 2nd free throw probabilities. It also simplifies the statistical assessment.

The next table reports the expected values of these statistics assuming $\Psi_1$, along with observed values for the NBA data. To avoid division by zero in any of the statistics, I only included players for whom there was at least one occurrence each of the four possible outcomes in a 2+ shot trip to the line, reducing the number of players in the sample from 1233 to 992.

| $\phi$ | $E(\phi|\Psi_1)$ | Average | StdErr | $z$ | Wtd Avg | Wtd StdErr | Wtd $z$ |
|---|---|---|---|---|---|---|---|
| $\hat{R}_i$ | 0.0059 | 0.0051 | 0.0009 | 6.980 | 0.0040 | 0.0003 | 15.315 |
| $R_i$ | 0.031 | 0.027 | 0.004 | 8.116 | 0.026 | 0.002 | 16.444 |
| $CD_i$ | 2.93% | 2.47% | 0.40% | 7.412 | 2.18% | 0.15% | 14.408 |

Table A4
Correlation Statistics
992 players, NBA 2000-2001 through 2013-2014 seasons

I computed *information-weighted* values (last three columns of Table A4), i.e., I used the reciprocal of the sample statistic's variance as the weight. Players with more trips to the free throw line receive more weight.

The uniformly-weighted averages are fairly close to $E(\phi|\Psi_1)$. However, the information-weighted averages are lower, and further from $E(\phi|\Psi_1)$. This is surprising in the context of performance variation at different time scales (see first paragraph, 'Possibilities for Future Research' section). Define

$$\hat{\Sigma}_i(t,\Delta t) = E\left[\Sigma_i | \Psi_1, \mathbf{Y}_i(t,\Delta t)\right]$$

$\mathbf{Y}_i(t,\Delta t)$: observations of trips by player $i$ between times $t$ and $t + \Delta t$

We would expect $\hat{\Sigma}_i(t,\Delta t)$ to increase with respect to $\Delta t$ if the dynamic process governing player $i$'s free throw shooting ability over time follows a random walk. If, for example, there were a systematic decline in free throw shooting at career's end, this would further increase $\hat{\Sigma}_i(t,\Delta t)$.

Possible explanations of this anomaly include:

(1) Players with more stable free throw shooting ability tend to have longer careers.
(2) Error correcting autoregression component in the dynamic process governing individual free throw performance.

These are not mutually exclusive hypotheses. They could even be mutually reinforcing. Error correcting autoregression can be generated by putting in more practice, or perhaps just by applying more mental focus, when a player misses more free throws than usual. For example, players who are more diligent about this, or with greater aptitude for correcting flaws, might tend to have longer careers.